\documentclass[twocolumn,showpacs,preprintnumbers,amsmath,amssymb,floatfix]{revtex4}
\usepackage{epsfig}

\usepackage{graphicx}
\usepackage{dcolumn}
\usepackage{bm}

\def\no{\nonumber}

\def\g{\gamma}

\def\s{\sigma}
\def\be{\begin{equation}}
\def\bea{\begin{eqnarray}}
\def\eea{\end{eqnarray}}
\def\ee{\end{equation}}
\def\bi{\begin{itemize}}
\def\ei{\end{itemize}}

\def\d{\delta}

\def\O{\Omega}

\def\vt{\vartheta}

\begin{document}

\title{Search templates for stochastic gravitational-wave backgrounds}

\author{Sukanta Bose}
\email{sukanta@wsu.edu}
\affiliation{%
Department of Physics and Astronomy, 
Washington State University, 1245 Webster, PO Box 642814, Pullman,
WA 99164-2814, U.S.A.
}%

\begin{abstract}
Several earth-based gravitational-wave (GW) detectors are actively pursuing 
the quest for placing observational constraints on models that predict 
the behavior of a variety of astrophysical and cosmological sources. These 
sources span a wide gamut, ranging from hydrodynamic instabilities in neutron 
stars (such as r-modes) to particle production in the early universe.
Signals from a subset of these sources are expected to appear in these 
detectors as stochastic GW backgrounds (SGWBs). The detection of these 
backgrounds will help us in characterizing their sources.
Accounting for such a background will also be required by some detectors, 
such as the proposed space-based detector LISA,
so that they can detect other GW signals. 
Here, we formulate the problem of constructing a bank
of search templates that discretely span the parameter space of a generic
SGWB. We apply it to the specific case of a class of cosmological SGWBs, known
as the broken power-law models. We derive how the template density varies in
their three-dimensional parameter space and show that for the LIGO 4km detector pair, 
with LIGO-I sensitivities, about a few hundred templates
will suffice to detect such a background
while incurring a loss in signal-to-noise ratio of no more than 3\%.

\end{abstract}


\maketitle

Multiple earth-based gravitational-wave (GW) detectors, including both 
resonant-mass 
and interferometric ones, are currently in operation aiming to make the 
first GW detection. As the sensitivities of these 
detectors improve, they will place interesting limits on astrophysical 
event rates and strengths of GW backgrounds, thus constraining or falsifying
theoretical models. The subject of this paper is how to design a
template bank for searching and bounding the strength of a
stochastic GW background (SGWB). After formulating the problem for a general 
SGWB, of either astrophysical or cosmological origin, we apply it
to the specific case of a 
SGWB with a spectral profile that belongs to a class 
predicted by a host of cosmological models, including inflationary and 
string-theoretic ones. This profile is known as the broken-power-law (BPL)
spectrum, as described below \cite{Maggiore:1999vm,Ungarelli:2003ty}. 

Searching for SGWBs with BPL-type spectra is important because some of the 
cosmological models that predict them also allow for their strengths to be 
large enough to be detectable in the near future.
In particular, in the bandwidth of these detectors, their strengths can be 
several orders of magnitude higher than that 
predicted by the slow-roll inflationary model (SRIM) \cite{Maggiore:1999vm}, 
while being consistent with
extant observational constraints, such as arising from the anisotropy in the 
cosmic microwave background radiation \cite{Allen:1994xz,Bose:2000}, 
the monitoring of radio pulses from several stable millisecond pulsars 
\cite{Kaspi:1994hp}, and the empirical abundances of light elements in the universe
\cite{kolbTurner90}. These models are, therefore, especially
attractive to the detectors in operation. Additionally, 
GWs from those unresolved astrophysical sources that have a duty cycle
appreciably larger than unity  will appear as a stochastic signal in
these detectors, some with BPL-type spectra and 
strengths that may be considerably larger than that
of a cosmological SGWB predicted by SRIM \cite{Ferrari:1998jf}. 
Indeed in the planned space-based detector LISA \cite{SysTech:2000rep},
the background from unresolved 
galactic binaries is expected to be large enough to 
form a ``source-confusion'' noise and make it difficult to detect other 
GW signals \cite{Bender:1997hs}. Although optimal statistics and templates
for searching resolvable binaries exist \cite{Krolak:2004xp,Rogan:2004wq}, 
strategies for characterizing such a background in order to allow the 
detection of other sources are beginning to be explored.
While we derive the limits on template spacings and template numbers
for BPL-type {\em cosmological} spectra, 
the formalism given here for obtaining the search templates is general enough 
to be applicable to other cases, including astrophysical SGWBs. 

We begin by briefly outlining the search statistic for a SGWB. 
The decision on whether a signal is present or absent in a detector output
is often based on the examination of data that are noisy.
In decision theory, the hypothesis that the data do not contain a signal is 
called the null hypothesis, $H_0$. Under the alternative hypothesis, $H_1$, 
the detector output is noise plus signal
\cite{Helstrom}. Thus, in the frequency domain, the output is 
\be\label{trueStrainH1}
\tilde{h}(f;\mbox{\boldmath $\vt$})|_{H_1}
= \tilde{n}(f) + {\cal A}\tilde{w} (f;\mbox{\boldmath $\vt$}) \ \ ,
\ee
where $\tilde{n}(f)$ is the noise, ${\cal A}$ 
is the overall {\em strength} of
the signal and  $\tilde{w} (f;\mbox{\boldmath $\vt$})$ 
gives the spectral characteristics of the signal for different choices of 
the {\em signal} parameter vector, $\mbox{\boldmath $\vt$}$.
In general, the waveform will have the appearance:
\be\label{waveform}
\tilde{w} (f; \mbox{\boldmath $\vt$}) 
= y(f; \mbox{\boldmath $\vt$})
\exp\left[i\Psi(f;\mbox{\boldmath $\vt$})\right] \ \ ,
\ee
where $y(f; \mbox{\boldmath $\vt$})$ is a frequency-dependent amplitude, and 
$\Psi(f;\mbox{\boldmath $\vt$})$ is the signal phase.
We assume that the detector noise has a zero-mean Gaussian probability
distribution; it is described completely by the first two noise moments,
\be
\overline{\tilde{n}(f)}=0 \quad {\rm and} \quad
\overline{\tilde{n}^*(f)\tilde{n}(f')}=\frac{1}{2}P(|f|)\delta(f-f') \ \ ,
\ee
where $P(|f|)$ is the one-sided noise power-spectral density (PSD)
\cite{saulsonBook}.
We define the inner (or scalar) product of
a pair of Fourier domain functions $\tilde{a}(f)$ and $\tilde{b}(f)$ as
\be \label{innerProd}
\left( \tilde{a} ,\>\tilde{b} \right) = 4 \Re \int_{0}^{\infty} \! df\>
\frac{\tilde{a}^* (f) \tilde{b}(f)}{m(f)}\ \ , 
\ee
where $\tilde{a}(f)$ and $\tilde{b}(f)$ are the Fourier
transforms of temporal counterparts $a(t)$ and $b(t)$, respectively, 
and $m$ is an inverse weight that, typically, depends on the noise PSDs.
Its exact form is decided by the detection statistic at hand.

The search templates are modeled after the waveform:
\be\label{template}
\tilde{u} (f; \mbox{\boldmath $\vt$}') 
= {\cal N}~\tilde{w}(f; \mbox{\boldmath $\vt$}')\ \ ,
\ee
where $\mbox{\boldmath $\vt$}'$ is the template parameter vector and
${\cal N}$ is a normalization factor. The inner-product of a template
with itself,
\be\label{norml}
\left( \tilde{u}(\mbox{\boldmath $\vt$}'),\>\tilde{u}(\mbox{\boldmath $\vt$}') 
\right) \equiv \kappa(\mbox{\boldmath $\vt$}') \ \ ,
\ee
will be taken to be positive definite. Above, $\sqrt{\kappa}$ is the template 
norm, which is parameter dependent, in general. The normalization factor is
related to the template norm as follows:
\be\label{normalization}
{\cal N}(\mbox{\boldmath $\vt$}')  = \frac{\sqrt{\kappa(\mbox{\boldmath $\vt$}')}}{2}\left[\int_0^\infty df\frac{y^2(f; \mbox{\boldmath $\vt$}')}
{m(f)}\right]^{-1/2} \,.
\ee
To test a hypothesis, one computes the cross-correlation of the data with the 
templates, {\it viz.}, $\left(\tilde{h}(\mbox{\boldmath $\vt$})|_{H_1},
\>\tilde{u}(\mbox{\boldmath $\vt$}')\right)$, which is termed as the
matched-filter output (MFO). Under $H_1$, the mean of the MFO is
\be\label{filterOutputGen}
\overline{S(\mbox{\boldmath $\vt$}, \d\mbox{\boldmath $\vt$})} :=
\overline{\left( \tilde{h}(\mbox{\boldmath $\vt$})|_{H_1},\>\tilde{u}(\mbox{\boldmath $\vt$}') \right)} \ \ ,
\ee
where  $\d \mbox{\boldmath $\vt$} \equiv 
\mbox{\boldmath $\vt$} - \mbox{\boldmath $\vt$}'$.

One often uses in searches unit-norm templates, namely,
\be\label{norml}
\kappa(\mbox{\boldmath $\vt$}')  =1 \,.
\ee
The advantage of using such templates is that under $H_1$ and for 
$\d \mbox{\boldmath $\vt$}  = 0$, the mean of the
matched-filter output (MFO) is just the 
signal strength divided by the template normalization factor, i.e.,
\be\label{filterOutputStrength}
\overline{S(\mbox{\boldmath $\vt$}, \d\mbox{\boldmath $\vt$}=0)} :=
\overline{\left( \tilde{h}(\mbox{\boldmath $\vt$})|_{H_1},\>\tilde{u}(\mbox{\boldmath $\vt$}) \right)} =\frac{\cal A}{{\cal N(\mbox{\boldmath $\vt$})}} \ \ ,
\ee
assuming that the signal model is perfect.

To quantify the effect of a mismatch, $\d \mbox{\boldmath $\vt$}$,
it is useful to introduce the match or ambiguity function:
\be\label{ambiguity}
M(\mbox{\boldmath $\vt$},\delta\mbox{\boldmath $\vt$})
:=\left( \tilde{u}(\mbox{\boldmath $\vt$}),\>\tilde{u}(\mbox{\boldmath $\vt$}') 
\right) \ \ ,
\ee
which tends to $\kappa(\mbox{\boldmath $\vt$})$ as $\delta\mbox{\boldmath $\vt$}\to 0$.
Then the mean of the MFO can be shown from Eq. (\ref{filterOutputGen}) to be
\be\label{meanSt1}
\overline{S(\mbox{\boldmath $\vt$},\delta\mbox{\boldmath $\vt$})} 
= \frac{\cal A}{{\cal N}(\mbox{\boldmath $\vt$})}M(\mbox{\boldmath $\vt$},\delta\mbox{\boldmath $\vt$}) \,.
\ee
For small values of $\d\mbox{\boldmath $\vt$}$, one can Taylor expand
$M$ about $\d\mbox{\boldmath $\vt$}=0$ to obtain
\bea\label{meanSt}
M (\mbox{\boldmath $\vt$}, \d\mbox{\boldmath $\vt$})
&=&\kappa\left(\mbox{\boldmath $\vt$}\right)
\Big[1 + d_{\mu}(\mbox{\boldmath $\vt$})\delta\vartheta^\mu \no\\
&&\quad\quad -g_{\mu\nu}(\mbox{\boldmath $\vt$})\delta\vartheta^\mu\delta\vartheta^\nu \Big]
+ {\cal O}\left(\delta\mbox{\boldmath $\vt$}^3\right)
\eea
where the Einstein summation convention over repeated indices,
$\mu$ and $\nu$, was used and we defined
\be\label{metric}
d_{\mu} := \frac{1}{\kappa}\left[\frac{\partial M}{\partial\delta
\vartheta^\mu}\right]_{\mbox{\boldmath $\d\vt$=0}} \,,
\>\>\>g_{\mu\nu} := -\frac{1}{2\kappa}\left[\frac{\partial^2 M}{\partial\delta
\vartheta^\mu\partial\delta\vartheta^\nu}\right]_{\mbox{\boldmath $\d\vt$=0}}
\,.\ee
Above, $g_{\mu\nu}$ can be interpreted as the metric on the parameter space
that maps parameter mismatches into dips in the signal-to-noise ratio (SNR)
\cite{Owen96}, provided $d_\mu$ vanishes. 
(The MFO 
of a unit-norm template is equivalent to the SNR 
\cite{boseCalib}.)

It is important to note here that an observer also has the choice of using
unnormalized templates, such that ${\cal N}=1$ in Eq. (\ref{template}).
This has the advantage that one does not have to recompute ${\cal N}$
and, therefore, the search templates, for every value of 
$\mbox{\boldmath $\vt$}$ or every time $m$ (which can be the
noise PSDs of the detectors)
changes. Indeed, this choice was exercised by Ungarelli and Vecchio in their
pioneering work in Ref. \cite{Ungarelli:2003ty}. However, the disadvantage of 
such a choice is that the associated ambiguity function,
$M (\mbox{\boldmath $\vt$}, \d\mbox{\boldmath $\vt$})$,
has first order errors arising from parameter mismatches. Consequently, 
a ``wrong'' template (i.e., a template with $\d\mbox{\boldmath $\vt$}\neq 0$)
applied to a given data set can actually trigger an MFO that is larger than 
that of the ``correct'' template (with $\d\mbox{\boldmath $\vt$}= 0$) applied
on the same data set.
Use of constant-norm templates, such as the unit-norm ones defined above, 
avoids that problem.

To see explicitly why $d_\mu$  need not be zero for unnormalized templates,
note that the associated ambiguity function obeys the Cauchy-Schwarz inequality
\cite{dennery}:
\be\label{Cauchy}
M (\mbox{\boldmath $\vt$}, \d\mbox{\boldmath $\vt$})=
\left( \tilde{u}(\mbox{\boldmath $\vt$}),\> \tilde{u}(\mbox{\boldmath $\vt$}')\right)
\leq \sqrt{\kappa(\mbox{\boldmath $\vt$})\kappa(\mbox{\boldmath $\vt$}')}
\,.\ee
Since, in general, $\kappa(\mbox{\boldmath $\vt$}')$ can be larger than 
$\kappa(\mbox{\boldmath $\vt$})$ for some $\mbox{\boldmath $\vt$}' \neq
\mbox{\boldmath $\vt$}$, the rhs above can actually exceed 
$\kappa(\mbox{\boldmath $\vt$})$. Thus, $\kappa(\mbox{\boldmath $\vt$})$
need not be the maximum value of $M (\mbox{\boldmath $\vt$}, \d\mbox{\boldmath $\vt$})$ for unnormalized templates. Equation (\ref{meanSt}) then implies that 
$d_\mu$  need not be zero for such templates. However, for unit-norm templates
the rhs of Eq. (\ref{Cauchy}) is identically unity (and 
$\kappa(\mbox{\boldmath $\vt$})=1=\kappa(\mbox{\boldmath $\vt$}')$), 
independent of the value 
of $\mbox{\boldmath $\vt$}'$ or $\mbox{\boldmath $\vt$}$. There, 
$M (\mbox{\boldmath $\vt$}, \d\mbox{\boldmath $\vt$})$ attains the maximum
possible value of 1, when
$\mbox{\boldmath $\vt$}'= \mbox{\boldmath $\vt$}$. Thus,
$d_\mu$ has to vanish for unit-norm (and constant-norm) 
templates, and $g_{\mu\nu}$ can assume its role as a parameter-space map.

Without any means for distinguishing a stochastic GW background in a 
detector from
the detector's intrinsic noise, the search for such a signal involves 
cross-correlating the outputs of a pair of detectors. As shown in Ref.
\cite{Allen:1997ad}, a useful statistic in decision making in this context
is the cross-correlation (CC) statistic,
\be\label{sgwbCC}
S:=\int_{-\infty}^{\infty}df\int_{-\infty}^{\infty}df'\d_T(f'-f)
\tilde{h}_{A}^*(f')\tilde{h}_{B}(f)\tilde{u}(f) \ \ ,
\ee
where $\tilde{h}_{A}(f)$ is the inverse Fourier transform of the strain in the
$A$th detector, $T$ is the observation time, $\d_T(f)\equiv \int_{-T/2}^{T/2}
dt~\exp(-i2\pi ft)$ is the finite-time 
approximation of the Dirac delta function, and $\tilde{u}(f)$ is a 
filtering function that will be determined below. 
The CC statistic can also be cast as the output of a matched filter:
\be\label{sgwbCCmatch}
S = \left( \tilde{K}|_{H_1},
\>\tilde{u} \right) \ \ ,
\ee
where $\tilde{K}|_{H_1}$ is a functional of a pair of detector inputs:
\be
\tilde{K}(f) := \frac{1}{2}\int_{-\infty}^{\infty}df'
\frac{\tilde{h}_{A}(f')\tilde{h}_{B}^*(f)\d_T(f'-f)}{P_{A}(f)P_{B}(f)}
\ee
and the inner product is defined as in Eq. (\ref{innerProd}), with the 
inverse weight there set to $m = \left[P_{A}(f)P_{B}(f)\right]^{-1}$.

The product $\left[\tilde{h}_{A}^*(f')\times\tilde{h}_{B}(f)\right]$ appearing 
in $S$ is a random variable since the SGWB strains, produced cosmologically or 
astrophysically (in some cases), are so. The detection statistic,
therefore, is the mean of the CC statistic,
\be\label{sgwbCCMean}
\overline{S} = \left( \overline{\tilde{K}|_{H_1}},\>\tilde{u} \right) \,.
\ee
And the variance of $S$ is
\be
\sigma^2=\overline{S^2}-
\left(\overline{S}\right)^2 \ \ ,
\ee
which defines the noise-squared of $S$. If one assumes 
that the noise power in each detector due to terrestrial sources is
much larger than that due to a SGWB, then \cite{Allen:1997ad}
\bea\label{CCVarianceTrue}
\sigma^2 &\approx& \overline{S^2} 
\approx \frac{T}{4}\int_{-\infty}^{\infty}df~\left|\tilde{u}(f)
\right|^2 ~P_{A}(|f|)P_{B}(|f|)\no\\
&=&\frac{T}{8} \left( \tilde{u},
\>\tilde{u} \right) \,.
\eea
Above, it was assumed in the second approximation that the cross-correlation 
of the terrestrial noises in the two detectors is negligible. Thus, the SNR is
\be\label{SNRUNTrueGeneral}
{\rm SNR} =\frac{\overline{S}}{\s}
= \sqrt{\frac{8}{T}}\frac{\left( \overline{\tilde{K}|_{H_1}},\>\tilde{u} \right)}
{\sqrt{\left( \tilde{u},\>\tilde{u} \right)}}
\ \ ,
\ee
which is maximized when the filtering function matches the signal, i.e.,
\be\label{CCtemplateGeneral}
\tilde{u} =\lambda \overline{\tilde{K}|_{H_1}} \,.
\ee
where $\lambda$ is a proportionality constant. Although $\overline{S}$ and 
$\s$ are dependent on the choice of this constant, the SNR itself is 
independent of it.

Equation (\ref{CCtemplateGeneral}) suggests 
$\tilde{u}$ as templates for searching an 
astrophysical or cosmological SGWB, as long as the assumptions made above remain
valid. We now concentrate in the rest of the paper on the search templates
required for a cosmological SGWB. (A similar problem for the astrophysical 
SGWB will be studied elsewhere \cite{boseASGWB}.)
Theoretically, the strain due to a cosmological SGWB
in each detector is expected to have a Gaussian probability 
distribution with zero 
mean; their variance-covariance matrix elements are
given by \cite{Christensen:1992wi}:
\bea\label{sgwbMeanTrue}
\overline{\tilde{h}_{A}^*(f)\Big|_{H_1}\tilde{h}_{B}(f')\Big|_{H_1}}
&=&\frac{3H_0^2}{20\pi^2}
|f|^{-3}\g_{AB}(|f|)\O_{\rm GW}(|f|;\mbox{\boldmath $\vt$})\no\\
&&\times\d(f-f') \ \ ,
\eea  
where $H_0$ is the Hubble constant, $\g_{AB}$ is the overlap-reduction 
function (ORF) for the detector pair \cite{Flanagan:1993ix}, 
$\O_{\rm GW}$ is the energy density of the stochastic GW per logarithmic 
frequency bin divided by the critical energy density required to close
the universe, and $\mbox{\boldmath $\vt$}$ are the signal parameters
on which it depends.
Heretofore, we will identify $\g_{AB}\equiv \g$. 
The ORF for co-located and co-aligned interferometric
detectors with orthogonal arms is normalized so that it is 
identically unity. Using
the above strain-power density in the expression for
$\overline{\tilde{K}|_{H_1}}$ yields the cosmological SGWB template:
\be\label{CCtemplate}
\tilde{u} ={\cal N}
\frac{\g(|f|)\O_{\rm GW}(|f|;\mbox{\boldmath $\vt$})}
{|f|^3 P_{A}(|f|)P_{B}(|f|)} \ \ ,
\ee
where, for unit-norm templates, 
\be\label{CCNorm}
{\cal N} = \left[2\int_{-\infty}^{\infty}df
\frac{\g^2(|f|)\O_{\rm GW}^2(|f|;\mbox{\boldmath $\vt$})}
{f^6 P_{A}(|f|)P_{B}(|f|)}\right]^{-1/2} \,.
\ee
We now show how to obtain the spacing between such templates on the 
$\mbox{\boldmath $\vt$}$ parameter space such that its discreteness, 
$\d \mbox{\boldmath $\vt$}$,
is small enough to guarantee an SNR of $\geq$97\% of that obtained in the
ideal case of  $\d \mbox{\boldmath $\vt$}=0$,

\subsubsection{Single power-laws}

Before studying the template spacings of the BPL spectra, let us consider
the simpler case where $\Omega_{\rm GW}(f)$ is in the form of a single
power-law (SPL) in frequency,
\be\label{powerLaw}
\Omega_{\rm GW}(f) = \Omega_0 \left(\frac{f}{f_p}\right)^k \ \ ,
\ee
where $k$ is a real power, and $\Omega_0$ and $f_p$ are positive-definite real 
constants. 
In the bandwidths of LISA or LIGO-type detectors, the SGWB spectrum predicted 
by SRIM will appear as a special case of the above, with $k\approx 0$. 
In such a case the only intrinsic 
search parameter is the index $k$ and the ambiguity function can be expanded 
around $\d k := k-k'=0$ as
\be\label{MpowerLaw}
M(k,\delta k) = 1 - \frac{1}{2}\left[\alpha(2)-\alpha^2(1)\right]\delta k^2 
+{\cal O}\left(\delta k^3 \right) \ \ ,
\ee
where $\alpha\left( n \right)$ is the template-space 
mean of $\ln \left(f/f_p\right)$:
\bea\label{alpha}
\alpha\left( n \right) &=& \left(\tilde{u}(k),\>
\left[\ln \left(f/f_p\right)\right]^n \tilde{u}(k)\right) \no\\
&=& \int_{f_-/f_p}^{f_+/f_p}dx \frac{\g^2(f_p x) x^{2k-6}
\left[\ln x\right]^n} {P_{A}(f_p x)P_{B}(f_p x)} \no\\
&&\times\left[\int_{f_-/f_p}^{f_+/f_p}dx \frac{\g^2(f_p x) x^{2k-6}}
{P_{A}(f_p x)P_{B}(f_p x)}\right]^{-1} \,.
\eea
Above, $n$ is any real number, and $f_-$, $f_+$ are the lower and upper 
limits of the frequency integral, respectively. For the LIGO-Hanford (LHO) 
and -Livingston (LLO) pair, with LIGO-I sensitivity, we choose
$f_- =40$Hz (determined by the detectors' seismic-noise cut-off) and 
$f_+ =512$Hz,
respectively. For this inter-site
correlation, the statistic does not receive any appreciable contribution
from higher frequencies owing to a weak $\gamma(f)$. By contrast, for 
co-located interferometers, $\gamma(f)=1$ for all frequencies, and the upper 
cut-off frequency is determined by the worsening sensitivity due to the 
photon shot-noise \cite{lazzariniWeiss,saulsonBook}. An optimal way of 
computing the CC statistic on the data from a set of multiple detectors that 
includes a co-located pair, while accounting for intra-site terrestrial noise,
was obtained in Ref. \cite{Lazzarini:2004hk}. For searches in this kind of a 
detector network, the value of $f_+$ should be revised upward of 512Hz.
As such, the expressions here are applicable to any pair of GW detectors. 
However, the template-spacings, number of 
templates, and the figures are computed for the LLO-LHO (4km) pair,
with LIGO-I sensitivities \cite{lazzariniWeiss}.
\\

\begin{figure}[!hbt] 
\centerline{\psfig{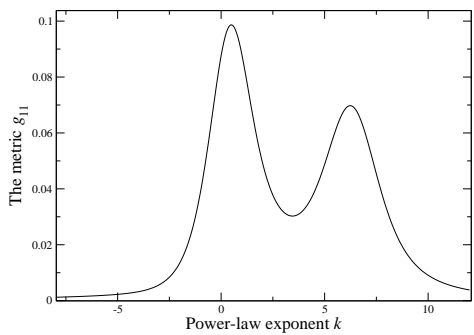}}
\caption{\small{The only metric component, $g_{11}$, for a single power-law SGWB 
plotted as a function of 
$k$. The template spacing 
is a minimum near the global maximum at $k = 0.5000^{+0.0625}_{-0.0625}$.
}}
\label{fig:spl}
\end{figure}

\begin{figure}[h]
\centerline{\psfig{file=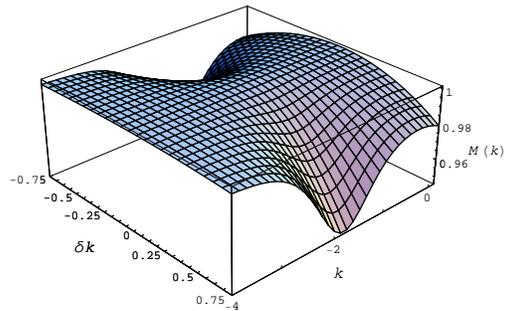,height=2.2in}}
\caption{\small{The ambiguity function, $M(k,\d k)$, for a single power-law 
SGWB plotted as a function of $k$ and the mismatch, 
$\d k$. For any given $k$, the function attains the maximum possible value 
of unity when $\d k=0$. And for any given $\d k$, the function is a minimum 
at $k= 0.5000^{+0.0625}_{-0.0625}$, which is consistent with the behavior of 
the metric depicted in Fig. \ref{fig:spl} \cite{numericalAmbiguityFunctions}. 
}}
\label{fig:splCrossCor}
\end{figure}

The only metric component for the above SGWB signal 
is readily deduced from Eq. (\ref{MpowerLaw}) as
\be\label{metricSPL}
g_{11}(k) = \frac{1}{2}\left[\alpha(2)-\alpha^2(1)\right] \ \ ,
\ee
which has the following properties: 
First, the Cauchy-Schwarz inequality can
be used to prove that $g_{11}(k)$ is non-negative for all real $k$
\cite{boseCalib}. Second, Eqs. (\ref{MpowerLaw}) and (\ref{alpha}) show that it is
dependent on $k$,
which is confirmed by Fig. \ref{fig:spl}.
This implies that for the optimal coverage of the $k$-space 
the template-spacings must be chosen to vary in step with the $g_{11}(k)$ values. 
The template spacing is a minimum at the global maximum of $g_{11}(k)$
(at 
$k \simeq 0.5$), and is:
\be
dk = 0.637\left(\frac{1-{\rm MM}}{0.01}\right)^{1/2} \ \ ,
\ee
where ${\rm MM}$ is the minimal match required by an observer between the 
discretely spaced templates and a signal. Typically, ${\rm MM}$ is set equal 
to 97\%. But we find above that $dk$ is as large as 0.637 for a minimal 
match as high as 99\%, as is evident in Fig. \ref{fig:splCrossCor}. 

If one were to choose the template-spacing to be uniformly equal to the above
value and, therefore, err on the side of over-covering the parameter space, then 
the number of templates required for a search with $k\in [-4,4]$ and
a minimal match of at least 99\% is \cite{kValues}
\be
N = {\rm Ceiling}\left[\frac{k_{\rm max} - k_{\rm min}}{0.637}\right] = 13 \ \ ,
\ee
which is easily implementable in real time on the data of a detector pair.

\subsubsection{Broken power-laws}

A likely character of $\Omega_{\rm GW}(f)$ will be in the form of a broken
power-law (BPL) in frequency \cite{Ungarelli:2003ty},
\be\label{BrokenPowerLaw}\small{
\Omega_{\rm GW}(f) = \Omega_0\Bigg[\Theta\left(1-\frac{f}{f_p}\right) 
\left(\frac{f}{f_p}\right)^{k_-}
+ \Theta\left(\frac{f}{f_p}-1\right) 
\left(\frac{f}{f_p}\right)^{k_+} \Bigg]\ \ ,}
\ee
where  $k_- \geq 0$, $k_+ \leq 0$ are real power-law exponents, and the
peak frequency, $f_p$, and $\Omega_0$ are positive-definite real constants. 
The first three are intrinsic search parameters, which define the three-dimensional 
parameter space, $\mbox{\boldmath $\vt$} = (f_p/f_{p0},k_-,k_+)$. Here $f_{p0}$ is an 
arbitrary reference frequency chosen large enough so that
$\d f_p/f_{p0}$ is small and the ${\cal O}\left(\delta\mbox{\boldmath $\vt$}^3\right)$ 
terms in Eq. (\ref{meanSt}) are indeed negligible.

We calculate the ambiguity function for the templates in Eq. (\ref{CCtemplate}),
with the above $\Omega_{\rm GW}(f)$, 
and compute the metric from its second derivatives using Eq. (\ref{metric}).
These derivatives now involve the following
integrals of the logarithm and different powers of $f$:
\bea\label{alphaBPL}
D &:=&\Bigg[\int_{f_-/f_p}^1 dx \frac{\g^2(f_p x) x^{2k_- -6}}
{P_{A}(f_p x)P_{B}(f_p x)}\no\\
&&\>\>+\int_1^{f_+/f_p} dx \frac{\g^2(f_p x) x^{2k_+ -6}}
{P_{A}(f_p x)P_{B}(f_p x)} \Bigg]^{-1} \ \ ,\no\\
\alpha_\pm\left( n \right) &:=& \mp D\int_{f_\pm/f_p}^1 dx \frac{\g^2(f_p x) x^{2k_\pm-6}
\left[\ln x\right]^n} {P_{A}(f_p x)P_{B}(f_p x)} \ \ , \no\\
\beta_\pm\left( n \right) &:=& \mp D k_\pm^n\int_{f_\pm/f_p}^1 dx \frac{\g^2(f_p x) x^{2k_\pm-6}} {P_{A}(f_p x)P_{B}(f_p x)} \ \ , \no\\
\beta(n) &:=& \beta_-(n) + \beta_+(n) \ \ ,
\eea
where the dependence of $D$, $\alpha_\pm(n)$, $\beta_\pm(n)$, and $\beta(n)$ on the
detector indices $A$ and $B$ is implicit.
\\

\begin{figure}[hbt] 
\centerline{\psfig{file=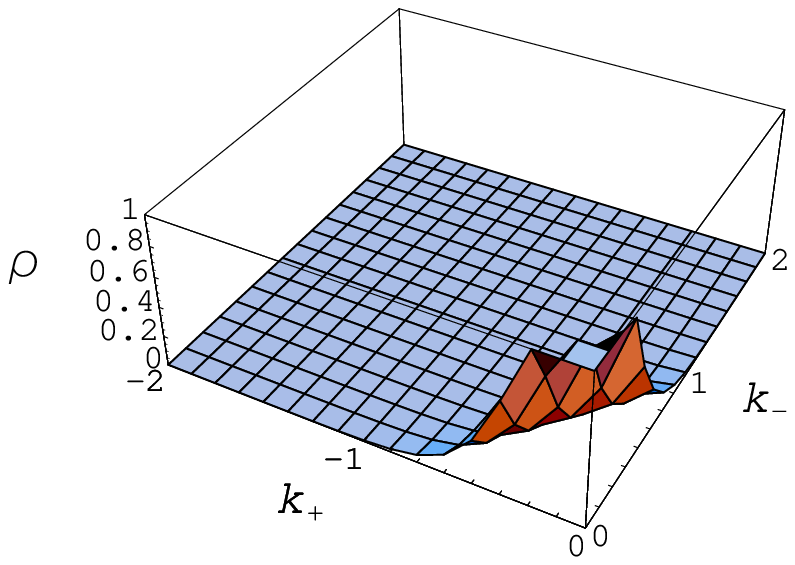,height=1.8in}}
\centerline{\psfig{file=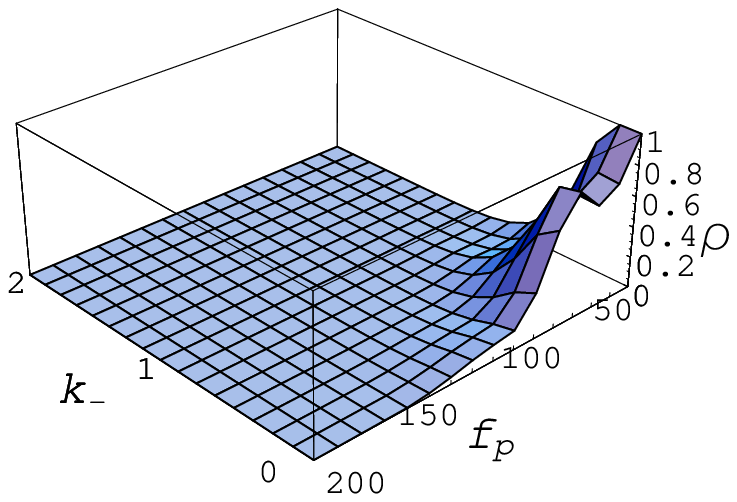,height=1.2in}
\psfig{file=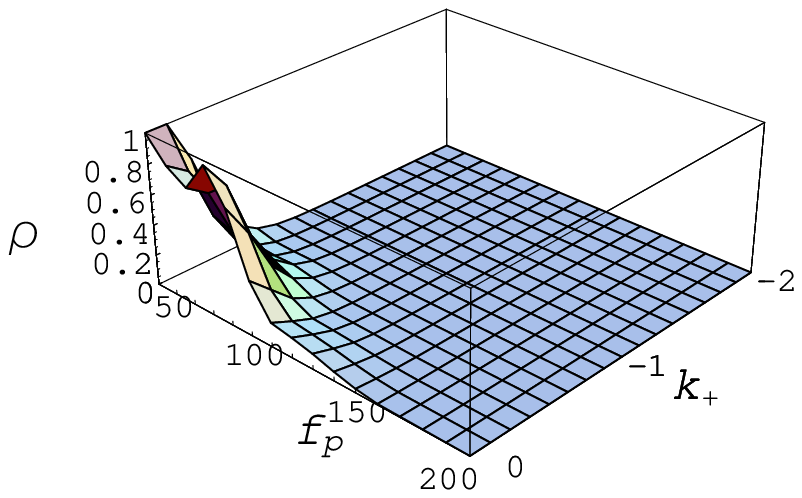,height=1.2in}}
\caption{
\small{The template-density $\rho$
(normalized here to have a maximum value of 1) for a BPL parameter-space 
plotted as a function of $k_\pm$ 
and $f_p$, two variables at a time.  The fixed parameters in these plots are 
(clockwise from top) $f_p=50$Hz, $k_- = 0.125$ and $k_+=-0.125$. 
}}
\label{fig:bplMetric}
\end{figure}

\begin{figure}[!hbt] 
\centerline{\psfig{file=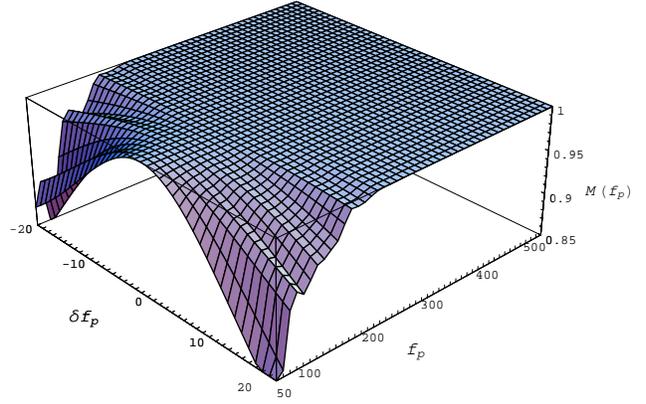,height=2.7in}}
\caption{\small{The ambiguity function, 
$M(\mbox{\boldmath $\vt$},\delta\mbox{\boldmath $\vt$})$, for a broken 
power-law SGWB plotted as a function of $f_p$ and the mismatch, $\d f_p$ 
(both in Hz). Here, we set $k_\pm = \mp 2$ and $\d k_\pm = 0$.
Note how quickly $M$ asymptotes to a value of unity for large $f_p$, 
implying a small number of templates for 
$f_p \stackrel{\Large{>}}{\small{\sim}} 200$Hz. The ambiguity function is 
within 99\% for: (a) $f_p \in [50, 120]$Hz, if $\d f_p \leq 4$Hz, 
(b) $f_p \in (120, 150]$Hz, if $\d f_p \leq 10$Hz, and 
(c) $f_p > 150$Hz, if $\d f_p \leq 20$Hz.
}}
\label{fig:bplCCfPeak}
\end{figure}

\begin{figure}
\centerline{\psfig{file=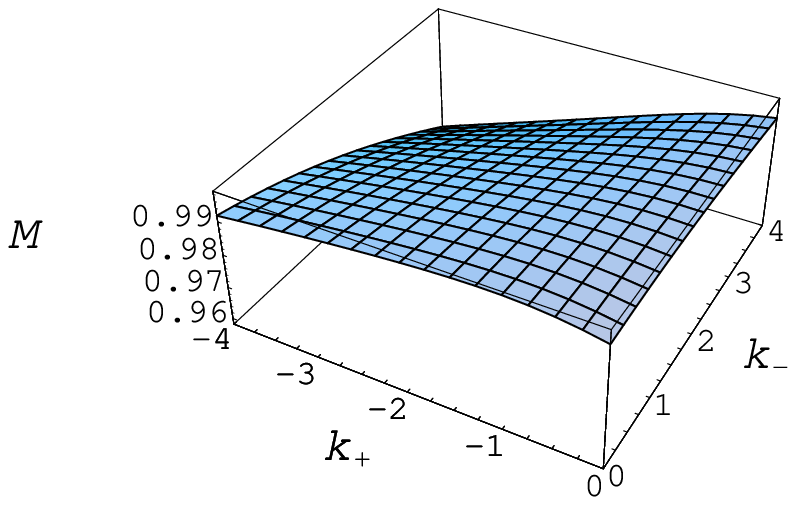,height=1.5in}
\psfig{file=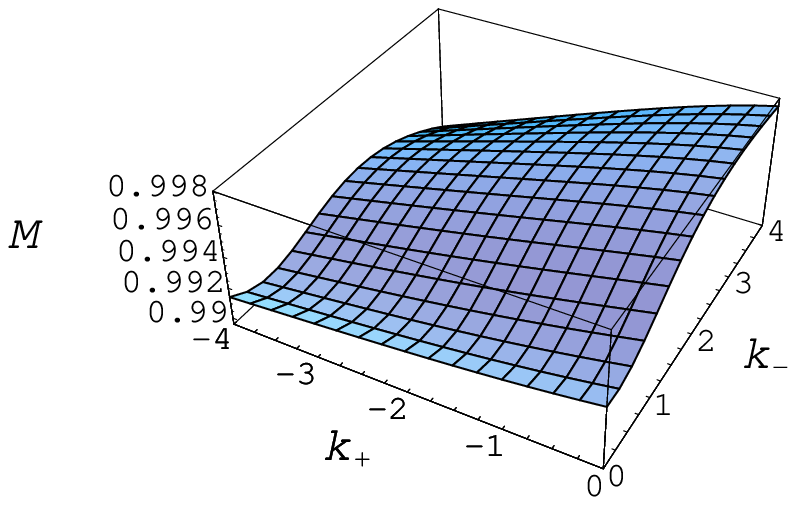,height=1.5in}}
\centerline{\psfig{file=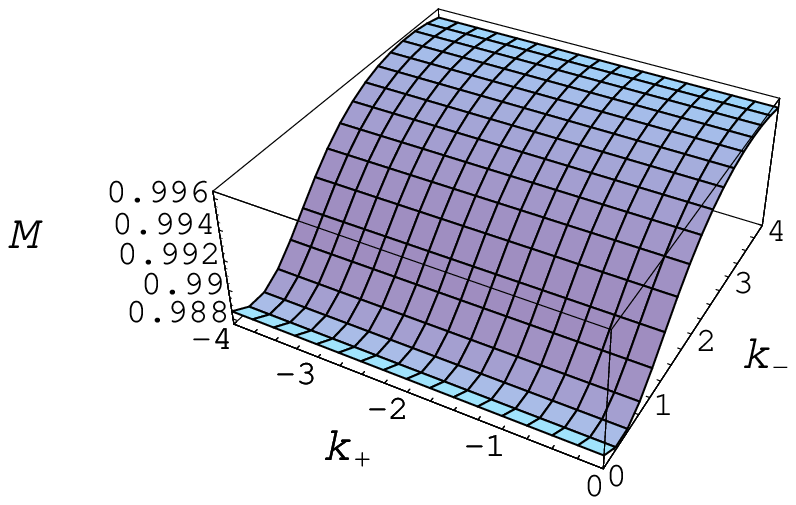,height=1.5in}
\psfig{file=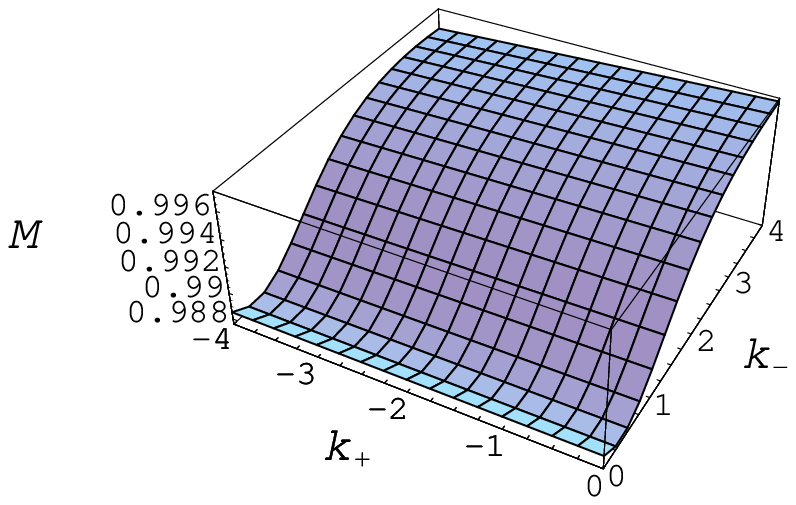,height=1.5in}}
\caption{
\small{$M(\mbox{\boldmath $\vt$},\delta\mbox{\boldmath $\vt$})$
for BPL spectra
plotted as a function of $k_+$ and $k_-$
for four values of $f_p=$ 50, 100, 190, 450Hz (clockwise from top left). 
The mismatch values are $\d k_\pm = 0.25$ and $\d f_p =2.5$Hz. As $f_p$
increases, note how little $M$ varies with $k_+$ (for any given $k_-$). 
This is indicative of the template spacings
being larger along $k_+$ than along $k_-$. This is expected since most 
of the contribution to the CC statistic arises at lower frequencies, where
the $k_-$ index has support (especially, when $f_p \stackrel{\Large{>}}{\small{\sim}} 200$Hz).
}}
\label{fig:bplMkpkm50to450}
\end{figure}

The six independent components of the symmetric 
metric on the three-dimensional parameter space, $(f_p/f_{p0},k_-,k_+)$, are:
\bea\label{metricBPL}
g_{00} &=& \frac{1}{2}\left(\frac{f_{p0}}{f_p}\right)^2\left[\beta(2)-\beta^2(1)\right]\no\\
g_{11} &=& \frac{1}{2}\left[\alpha_-(2)-\alpha_-^2(1)\right] \ \ ,\no\\
g_{22} &=& \frac{1}{2}\left[\alpha_+(2)-\alpha_+^2(1)\right]\ \ ,\no\\
g_{01} &=& \frac{1}{2}\left(\frac{f_{p0}}{f_p}\right)\left[\beta(1) - k_-\right]\alpha_-(1)\ \ ,\no\\
g_{02} &=& \frac{1}{2}\left(\frac{f_{p0}}{f_p}\right)\left[\beta(1) - k_+\right]\alpha_+(1)\ \ ,\no\\
g_{12} &=& -\frac{1}{2}\alpha_-(1)\alpha_+(1) \ \ ,
\eea
where the parameter vector components were taken to be 
$\vt^0 = f_p/f_{p0}$, $\vt^1 = k_-$, and $\vt^2 = k_+$.
Once again it is clear that the metric is dependent on the parameters, as
confirmed by Fig. \ref{fig:bplMetric}.
Thus, the template spacing for a fixed minimal match is not uniform 
\cite{inflateBPL}.

\begin{figure}[hbt] 
\centerline{\quad\quad\psfig{file=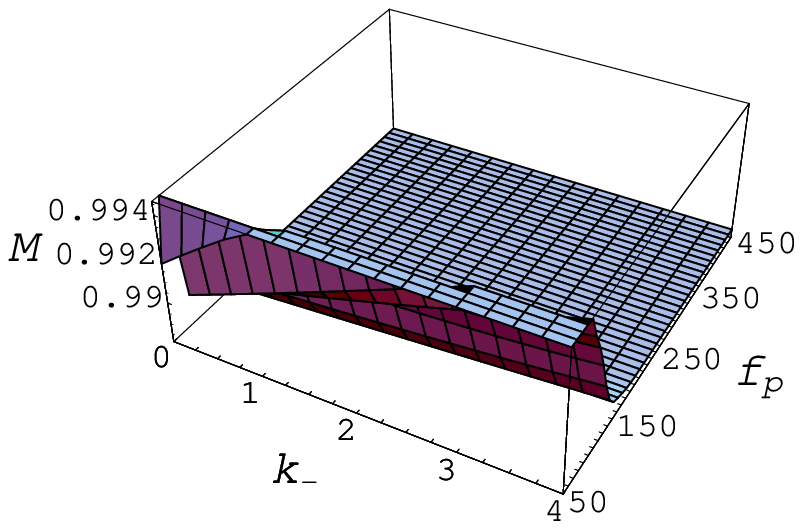,height=1.5in}
\psfig{file=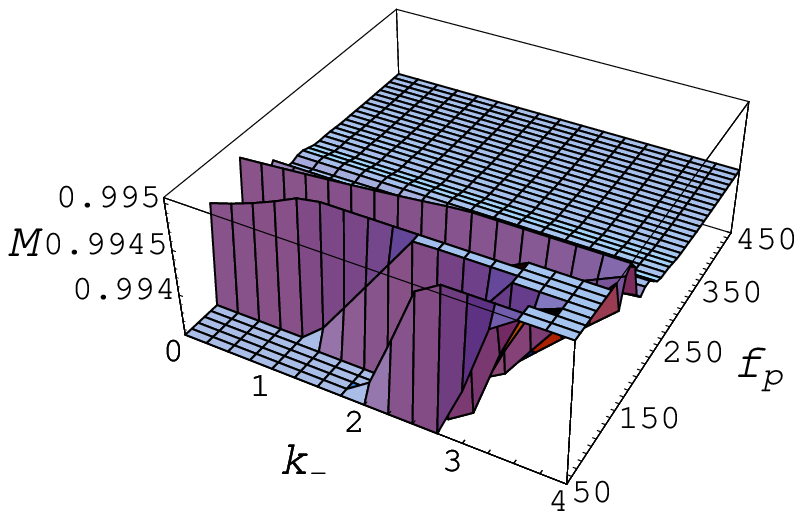,height=1.5in}}
\centerline{\quad\quad\psfig{file=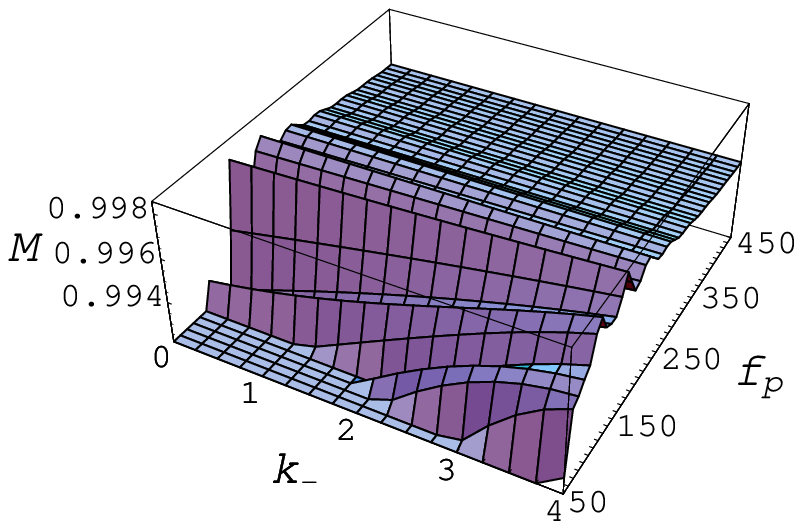,height=1.5in}
\psfig{file=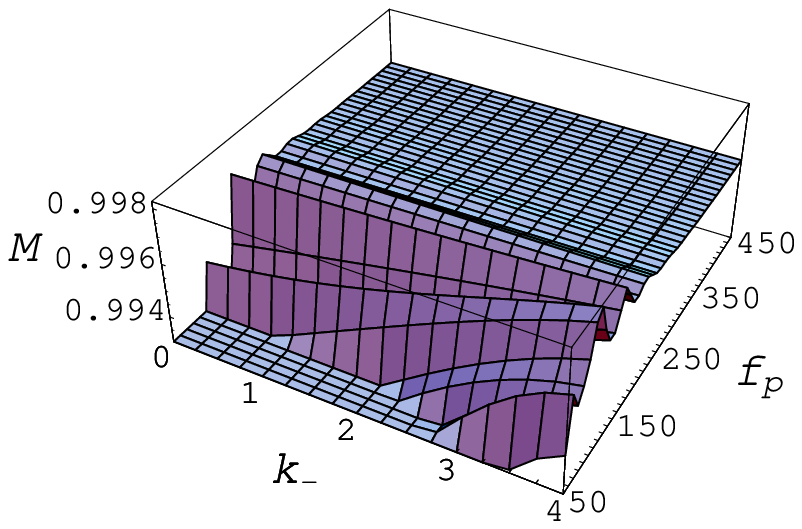,height=1.5in}}
\caption{
\small{$M(\mbox{\boldmath $\vt$},\delta\mbox{\boldmath $\vt$})$
for a broken power-law SGWB plotted as a function of $k_-$ and $f_p$
for four values of $k_+=$ -4, -2, -1, 0 (clockwise from top left). 
The mismatch values are $\d k_\pm = 0.25$ and $\d f_p =2.5$Hz. 
}}
\label{fig:bplMkmfp50to450}
\end{figure}

\begin{figure}[hbt] 
\centerline{\psfig{file=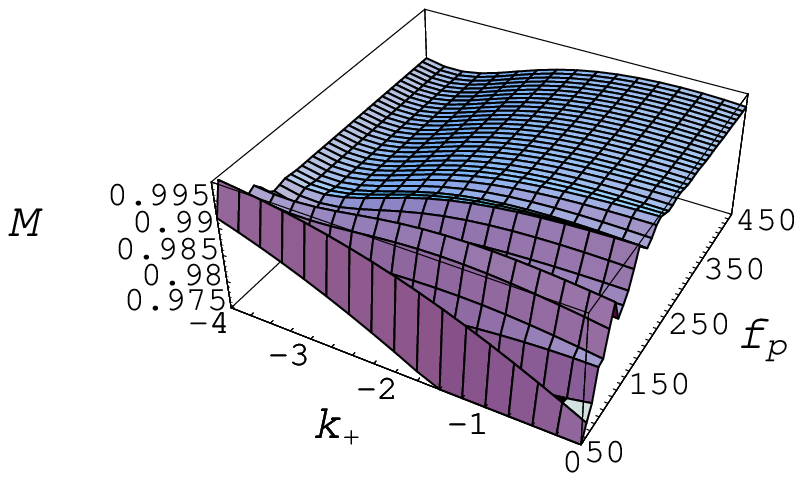,height=1.5in}
\psfig{file=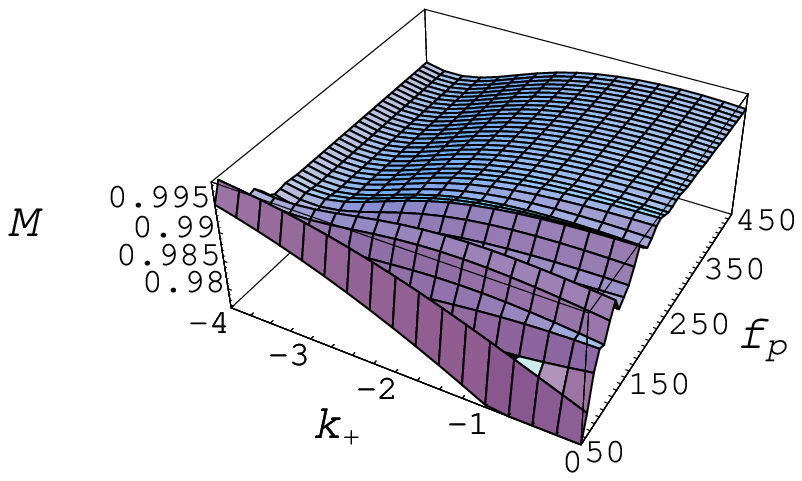,height=1.5in}}
\centerline{\psfig{file=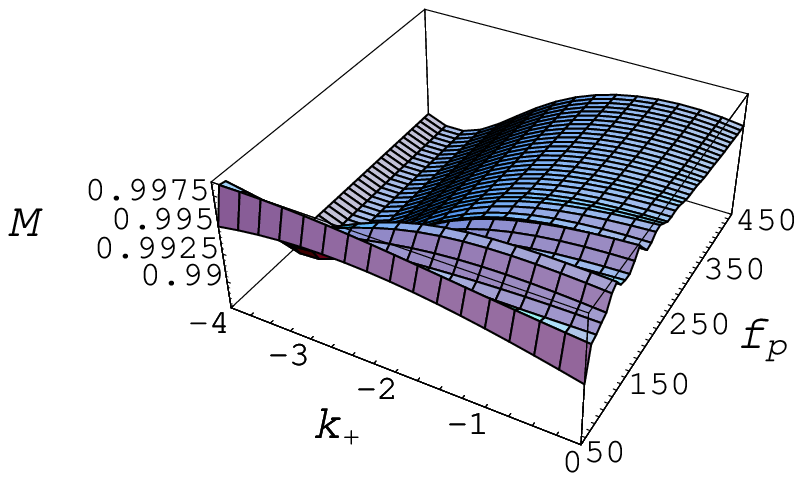,height=1.5in}
\psfig{file=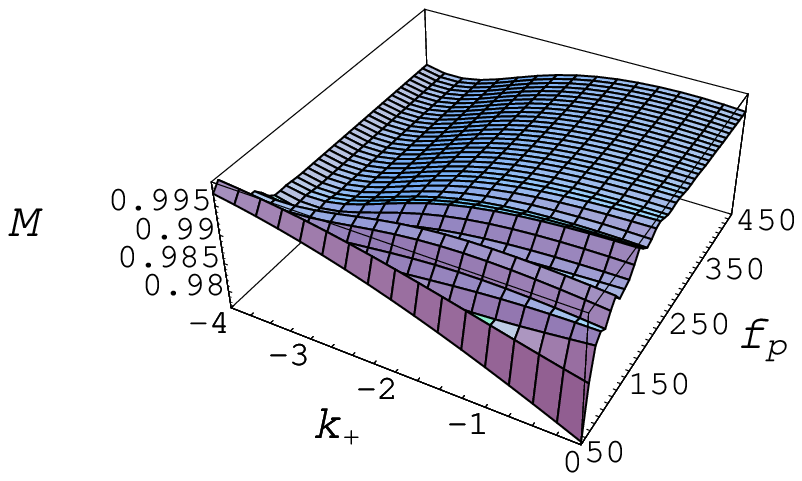,height=1.5in}}
\caption{
\small{$M(\mbox{\boldmath $\vt$},\delta\mbox{\boldmath $\vt$})$
for a broken power-law SGWB plotted as a function of $k_+$ and $f_p$
for four values of $k_-=$ 0, 1, 2, 4 (clockwise from top left). 
The mismatch values are $\d k_\pm = 0.25$ and $\d f_p =2.5$Hz. 
}}
\label{fig:bplMkpfp50to450}
\end{figure}

The number of templates, $N$, can be estimated from the above metric by 
dividing the parameter space volume, 
\be
V = \int d^3\vt \sqrt{\det \|g_{\mu\nu}\|} \ \ ,
\ee
by the volume of a unit cell in three-dimensional space, 
$v=(2\sqrt{(1-{\rm MM})/3}~)^3$ \cite{Owen96}. We numerically compute the above
volume and find that
\be
N \simeq 250 \left(\frac{1-{\rm MM}}{0.03}\right)^{-3/2} \ \ ,
\ee
where we again err on the side of over-coverage by relaxing the numerical 
computation to allow for ${\rm MM} \geq$ 97\%. The parameter-dependence of
the template density, $\rho = \sqrt{\det \|g_{\mu\nu}\|}/v$, is illustrated
in Fig. \ref{fig:bplMetric}. The above value of $N$ is consistent
with the near-unity value of the ambiguity function shown in Figs.
\ref{fig:bplCCfPeak}-\ref{fig:bplMkpfp50to450}
for template-spacings as large as $\d k_\pm = 0.25$ and $\d f_p =2.5$Hz.

It has been projected in Ref. \cite{Allen:1997ad} that LIGO-I and Advanced
LIGO may succeed in placing upper limits on $\Omega_0$ of the order of
$5\times 10^{-6}$ and $5\times 10^{-11}$, respectively,
for the $k=0$ SPL spectrum. The first science
run at LIGO already demonstrated successfully the application of a single
template (i.e., the $k=0$ case of SPL) on the data from the LIGO detector 
pairs to obtain bounds on $\Omega_0$ \cite{ligoS1,Bose:2003nb}. With the 
upcoming science runs at LIGO, the sensitivities are fast approaching closer to 
the designed target so as to make the first upper limit given above 
realizable in the near future. This progress 
necessitates the availability of techniques and templates to look for a variety 
of proposed astrophysical and cosmological SGWBs in the ever-so sensitive data. 
This paper addresses this issue for the latter category of signals, which
assumes the background to be isotropic and unpolarized. The former case of
an astrophysical background will be discussed elsewhere \cite{boseASGWB}.

\acknowledgments

I would like to thank Aaron Rogan for help in plotting some of the figures.
Thanks are also due to John Whelan for critically reviewing the manuscript 
and offering helpful comments and to Carlo Ungarelli for making useful 
suggestions. 
This research was funded in part by NSF Grant PHY-0239735
and NASA Grant NAG5-12837.


\begin{references}
\bibitem{numericalAmbiguityFunctions}
All plots of the ambiguity function in this paper are computed numerically and 
give a more accurate, but not appreciably different, value of $M(k,\d k)$ than 
what can be estimated from Eq. (\ref{meanSt}).
\bibitem{kValues}
The range of $k$ values that an observer is interested in searching for will 
depend on the availability of models predicting SGWBs with those values and 
the noise spectral profiles of the detector pair. The range chosen
here is for illustrative purposes. However, the metric expression given here
can be applied to estimate the required number of templates for a wider range of
 $k$ values.
\bibitem{inflateBPL}
Note that the slow-roll inflationary SGWB spectrum will appear in the bandwidth
of LISA- or LIGO-type detectors as a 
$k\approx 0$ SPL spectrum, and will be indistinguishable from a BPL spectrum 
if either $k_- \approx 0$ and 
$f_p > f_+$ or $k_+ \approx 0$ and $f_p < f_-$. In this limited sense, the SRIM spectrum
can be construed as a special case of the BPL.
\end{references}
\end{document}